\documentclass[journal=jacsat,manuscript=article]{achemso}
\SectionNumbersOn
\usepackage[version=3]{mhchem} 
\usepackage{amsmath}
\usepackage{mathrsfs}
\usepackage{caption}
\usepackage{subcaption}
\usepackage{epigraph}

\usepackage{hyperref}

\author{Marcel D. Fabian}
\email{marcel.fabian@nbi.ku.dk}
\affiliation[]{NNF Quantum Computing Programme, Niels Bohr Institute, University of Copenhagen, Denmark}
\author{Nina Glaser}
\affiliation[]{NNF Quantum Computing Programme, Niels Bohr Institute, University of Copenhagen, Denmark}
\alsoaffiliation[]{Nano-Science Center and Department of Chemistry, University of Copenhagen, Denmark}
\author{Gemma C. Solomon}
\affiliation[]{NNF Quantum Computing Programme, Niels Bohr Institute, University of Copenhagen, Denmark}
\alsoaffiliation[]{Nano-Science Center and Department of Chemistry, University of Copenhagen, Denmark}

\title[]
  {The PPP model - a minimal viable parametrisation of conjugated chemistry for modern computing applications}

\abbreviations{IR,NMR,UV}

\begin{document}
\newpage
%
%
%
%
%

\begin{abstract}
    The semi-empirical Pariser-Parr-Pople (PPP) Hamiltonian is reviewed for its ability to provide a minimal model of the chemistry of conjugated $\pi$-electron systems, and its current applications and limitations are discussed.
    From its inception, the PPP Hamiltonian has helped in the development of new computational approaches in instances where compute is constrained due to its inherent approximations that allow for an efficient representation and calculation of many systems of chemical and technological interest.
    The crucial influence of electron correlation on the validity of these approximations is discussed, and we review how PPP model exact calculations have enabled a deeper understanding of conjugated polymer systems.
    More recent usage of the PPP Hamiltonian includes its application in high-throughput screening activities to the inverse design problem, which we illustrate here for two specific fields of technological interest: singlet fission and singlet-triplet inverted energy gap molecules.
    Finally, we conjecture how utilizing the PPP model in quantum computing applications could be mutually beneficial. 
\end{abstract}

\section{Introduction}
\begin{quote}
  \textit{``The underlying physical laws necessary for the mathematical theory of a large part of physics and the whole of chemistry are thus completely known, and the difficulty is only that the exact application of these laws leads to equations much too complicated to be soluble.
  It therefore becomes desirable that approximate practical methods of applying quantum mechanics should be developed, which can lead to an explanation of the main features of complex atomic systems without too much computation.''} \\ P. A. M. Dirac, 1929
\end{quote}

The famous lines by Dirac from the introductory paragraph of his publication on quantum mechanics of many-electron systems\cite{dirac_quantum_1929} offer a fascinating glimpse into the early formative years of quantum mechanics. Optimism regarding the veracity of the formalism was met by the realisation of the daunting nature of solving any chemically relevant multi-electron system.
Before the invention of the digital computer, computation was a manual human endeavour and quantum mechanical description of many electron systems was beyond the scope of what was achievable. In the years that followed, theoretical chemistry has benefited tremendously from the introduction of computers and the steadily increasing hardware resources, following Moore's law, which eventually allowed the description of molecules and other systems of ever-increasing complexity.
Nonetheless, the nature of these “complex atomic systems” means that an exact description is still outside the realms of computational feasibility.

The Pariser-Parr-Pople (PPP) Hamiltonian was devised as such an “approximate practical method” for the electronic structure problem, and in this perspective we follow its historic development and explore the future prospects in modern computing.
We describe the approximations that have made the PPP Hamiltonian practical for chemistry, and why it remains relevant even as increasing computational capabilities removed earlier constraints.
Finally, we give an outlook on how the PPP Hamiltonian might be impactful for quantum computing, which currently faces similar resource constraints to those seen in classical computing 70 years ago.

\section{The origins of the PPP model}
\label{section::OriginsPPP}
\subsection{The Hückel model}
\label{subsec::Hueckel}
The origin of the PPP methodology traces back to the beginnings of quantum mechanics itself.
In 1931 Hückel formulated the molecular orbital (MO) theory for conjugated molecules \cite{huckel_quantentheoretische_1931-1,huckel_quantentheoretische_1931,huckel_quantentheoretische_1932,huckel_freien_1933}, now known as the Hückel MO (HMO) model.
It was later extended by Lennard-Jones, Coulson, and Longuet-Higgins to the general theory of $\pi$-electrons for unsaturated and aromatic molecules\cite{trinajstic_new_1977}.
With very limited computational effort, the HMO theory provided a qualitative understanding of $\pi$-electron conjugated systems and thus the ability for organic chemists to predict and plan experiments \cite{trinajstic_new_1977}.
At that time, this essentially meant calculations were done with pen and paper.
While its use in state-of-the-art computation today has greatly diminished, HMO theory remains very valuable in teaching chemical concepts such as the chemical bond or the MO-LCAO approach (\textbf{m}olecular \textbf{o}rbitals from \textbf{l}inear \textbf{c}ombinations of \textbf{a}tomic \textbf{o}rbitals) and is included in most chemistry textbooks\cite{kutzelnigg_what_2007}. 
\begin{figure}[htb]
    \centering
    \includegraphics[width=0.3\textwidth]{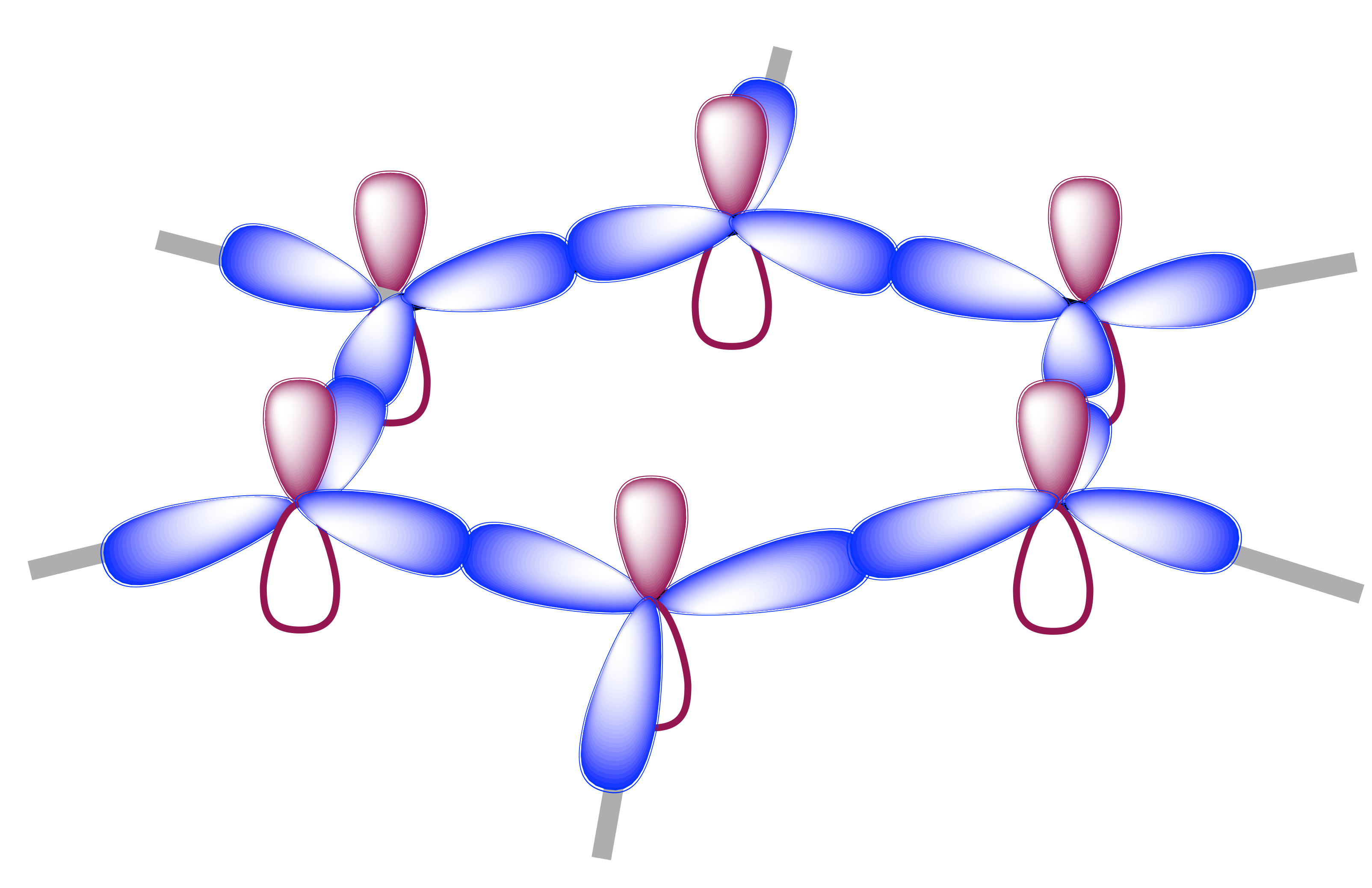}
    \caption{Schematic illustration of orbitals that can form  $\sigma$-bonds (blue) and $\pi$-bonds (purple) in benzene}
    \label{fig::SigmaPiBenzene}
\end{figure}

The HMO model relies on several assumptions\cite{kutzelnigg_what_2007} that are also relevant for our introduction to the PPP model, and we will summarize them here.
The most consequential assumption is the distinction between two different types of one-electron functions that are called $\sigma$- and \mbox{$\pi$-MOs}, respectively.
In the case of planar molecules, the \mbox{$\sigma$-MOs} are symmetric with respect to the molecular plane and are linear combinations of atomic orbitals (AOs) centred on the various atoms of the molecule. The \mbox{$\sigma$-MOs} define the structural backbone of the conjugated molecule, are assumed to be localized and transferable between the same atom types. Importantly, the \mbox{$\sigma$-MOs} and are not treated explicitly in HMO theory.

The $\pi$-MOs are conceptually linear combinations of $2 p_z$-orbitals that are antisymmetric with respect to the molecular plane (see Fig.~\ref{fig::SigmaPiBenzene}).
There are up to 2 electrons in each delocalized $\pi$-MO, that can move through the bonds between neighbouring atoms, defined by the resonance integral.
The overlap between $2p_z$-orbitals is assumed to be orthonormal in the zero differential overlap (ZDO) approximation \cite{lipkowitz_semiempirical_1991}.
This requires replacement of the original resonance integral, with a reduced resonance integral\cite{kutzelnigg_what_2007,mulliken_quelques_1949}, which is treated in HMO theory as a semi-empirical parameter.
Finally, the HMO model is a one-electron theory, which means that the electrons do not explicitly interact through the Hamiltonian.

The omission of electron-electron interactions leads to one of the fundamental shortcomings of HMO theory\cite{jug_theoretical_1990, pople_application_1957}. The electron-electron interaction terms were (and still can be) challenging to calculate: first, they are generally non-analytic, unless the AO basis is chosen to be represented by Gaussians, as suggested later by Boys\cite{boys_electronic_1950}.
Furthermore, when treated naively, the number of terms grows as $O(N^4)$, with $N$ being the number of orbitals. Due to these challenges, electron-electron interactions were typically not treated explicitly in quantum chemical calculations up to the early 1960s \cite{parr_quantum_1972} for anything but the smallest systems.
As anecdotal evidence of the constrained computational power, we highlight that up to 1960 a grand total of 80 full \textit{ab\,initio} calculations had been performed on molecules with 3 or more electrons \cite{parr_quantum_1972,allen_basis_1960}.
The first capable workstations and mainframes started to emerge at that time\cite{dewar_molecular_1969} and quickly grew in capabilities.
It was, however, clear that experimental chemists were interested in much larger molecules and their spectroscopic properties, which were inaccessible even with these growing computational resources.
Hence, there was a large need for an approximate treatment of electron-electron interaction before an explicit inclusion of these terms was computationally feasible.
This novel model would have to retain the simplicity and computability of the HMO theory, but through the inclusion of some form of electron-electron interaction, it would hopefully enable a more quantitative agreement with experimental results. 

\subsection{The PPP model}
\label{subsec::PPP}
Such an extension to the HMO theory was indeed proposed in 1953 by Pariser and Parr \cite{pariser_semi-empirical_1953,pariser_semi-empirical_1953-1}, and separately also by Pople\cite{pople_electron_1953}.  Due to the similarity of the underlying assumptions, these methods are unified as the Pariser-Parr-Pople (PPP) model. 
The key insight for both proposals was that the ZDO approximation (see section~\ref{subsubsec::ZDO}), already invoked in the HMO for the overlap matrix, would also simplify and reduce the number of electron-electron interaction integrals from $O(N^4)$ to a more manageable number of $O(N^2)$\cite{jug_theoretical_1990}.
These integrals were parametrized based on experimentally available data, which is why both the Hückel and PPP model belong to the class of semi-empirical approaches (see section~\ref{subsec::SemiEmpirical}).  

Pariser and Parr started from the Hückel MOs and performed a configuration interaction (CI) calculation \cite{pariser_semi-empirical_1953,pariser_semi-empirical_1953-1}.
This calculation could run on desk calculation machines available at the time\cite{dewar_molecular_1969} by virtue of the reduction of integral terms that had to be evaluated.
Pople also started from an initial Hückel guess and then self-consistently solved the Roothaan-Hall equation while invoking the ZDO approximation \cite{pople_electron_1953,pople_electronic_1955}.
The original suggestion of Pople was therefore tailored towards the ground state, whereas Pariser and Parr targeted the first few excited states \cite{dewar_molecular_1969}.
In later years, the combination of an initial self-consistent treatment with a subsequent limited CI calculation (such as e.g. CIS)\cite{pople_electronic_1955} became very popular by virtue of its success in describing both ground states and electronic excitation spectra of organic molecules.

\subsection{Beyond PPP: general semi-empirical and \texorpdfstring{\textit{ab\,initio}}{ab initio} methods}
\label{subsec::SemiEmpirical}
While the ZDO approximation greatly reduces the number of non-zero electron-electron interaction terms, the remaining integrals still have to be determined somehow.
Instead of calculating these integrals explicitly as in \textit{ab initio} methods, Pariser and Parr suggested a semi-empirical approach, where the integrals are treated as parameters and fitted to certain experimental values\cite{pariser_semi-empirical_1953,pariser_semi-empirical_1953-1}.
This calibration also helps compensate for neglected terms\cite{thiel_semiempirical_2014}.

The initial limitation of the ZDO approximation to planar $\pi$-electron systems was lifted when Pople introduced the so-called complete neglect of differential overlap (CNDO) and neglect of diatomic differential overlap (NDDO) approximations that could treat general three-dimensional systems with $\sigma$-electrons explicitly\cite{pople_approximate_1965}.
Based on these initial schemes, many semi-empirical methods have been proposed with increasing sophistication and generally also a larger number of parameters\cite{segal_neglect--differential-overlap_1977, lipkowitz_semiempirical_1990, lipkowitz_semiempirical_1991, thiel_semiempirical_1988, thiel_semiempirical_2005, thiel_semiempirical_2014}.
The parametrization for these methods followed different philosophies, with either fitting to experimental data or to \textit{ab\,initio} Hartree-Fock (HF) results and targeting either ground state or excited state properties \cite{murrell_semi-empirical_1972,segal_semiempirical_1977-1,thiel_semiempirical_2014}.

We want to mention the great contention over the correct parameterisation in specific semi-empirical methods \cite{clark_quo_2000,pople_deficiencies_1975,hehre_mindo3_1975,dewar_concerning_1975,dewar_applications_1985} here because it highlights the strengths and weaknesses of semi-empirical approaches in general and specifically in relation to \textit{ab initio} approaches.
Most semi-empirical models are built on the MO \textit{ab\,initio} framework, but through the neglect or parameterisation of certain integrals can treat much larger systems\cite{thiel_semiempirical_2014}.
This is reliable within the limitations of the parametrisation, but can fail drastically outside.
It is not necessarily clear \textit{a priori} how well the parameters transfer between different systems when no reference data are available.  
Here, we mention an early version of the MINDO semi-empirical method as an example where the parametrisation failed, resulting in large errors in the heats of atomization/formation for highly strained molecules\cite{dewar_molecular_1969}.

Broadly, semi-empirical treatments lack the generality of a fully \textit{ab\,initio} approach, where the path to improve results is known, namely through a higher level of theory or a larger basis sets.
At the same time, however, semi-empirical treatments might result in better agreement with experiments than \textit{ab\,initio} calculations, while also being much faster and applicable to larger systems when limited computational resources are available.
In terms of speed and generality, semi-empirical methods can be placed in-between a full \textit{ab\,initio} treatment and molecular mechanics (MM) models.

\section{PPP as the MVP of chemistry}
\label{section::MVPPPP}
The PPP Hamiltonian has been remarkably successful in computational chemistry applications, especially in the description and determination of excited-state properties of conjugated $\pi$-electron systems.
In the following section, we will highlight why the PPP model is the minimal viable parametrisation (MVP) to describe many chemically relevant systems through comparison with other model Hamiltonians.
We will then describe how many of the empirical approximations of the PPP Hamiltonian have been theoretically explained and validated once accurate \textit{ab\,initio} calculations became computationally feasible.

\subsection{Comparison with other models}
\label{subsec::comparison}
\subsubsection{Hückel Hamiltonian}
\label{subsubsec::Hueckelmodel}

\begin{figure*}[htb]
     \centering
     \begin{subfigure}[b]{0.23\textwidth}
         \centering
         \includegraphics[width=\textwidth]{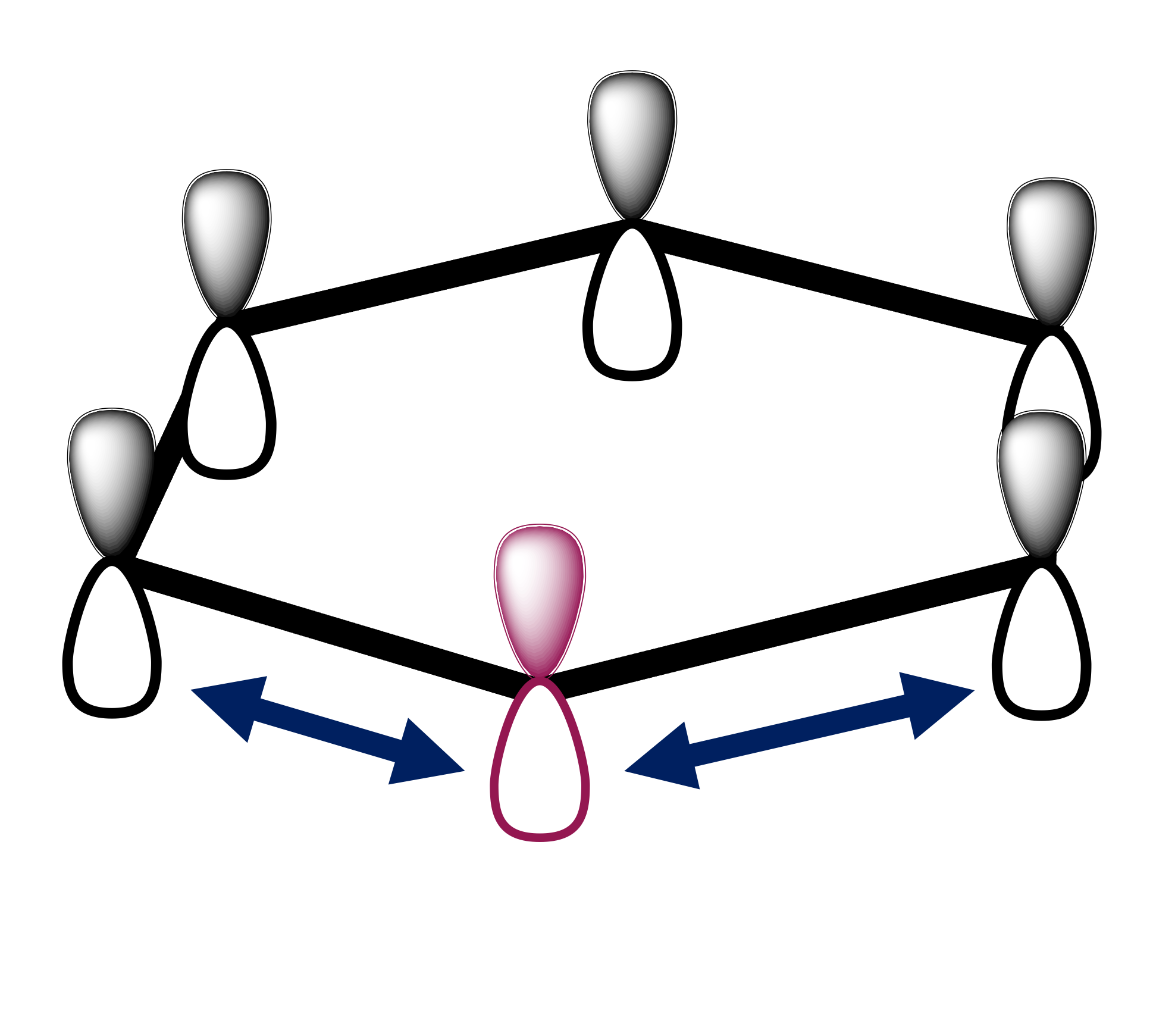}
         \caption{Hückel}
         \label{fig::Hueckel}
     \end{subfigure}
     \hfill
     \begin{subfigure}[b]{0.23\textwidth}
         \centering
         \includegraphics[width=\textwidth]{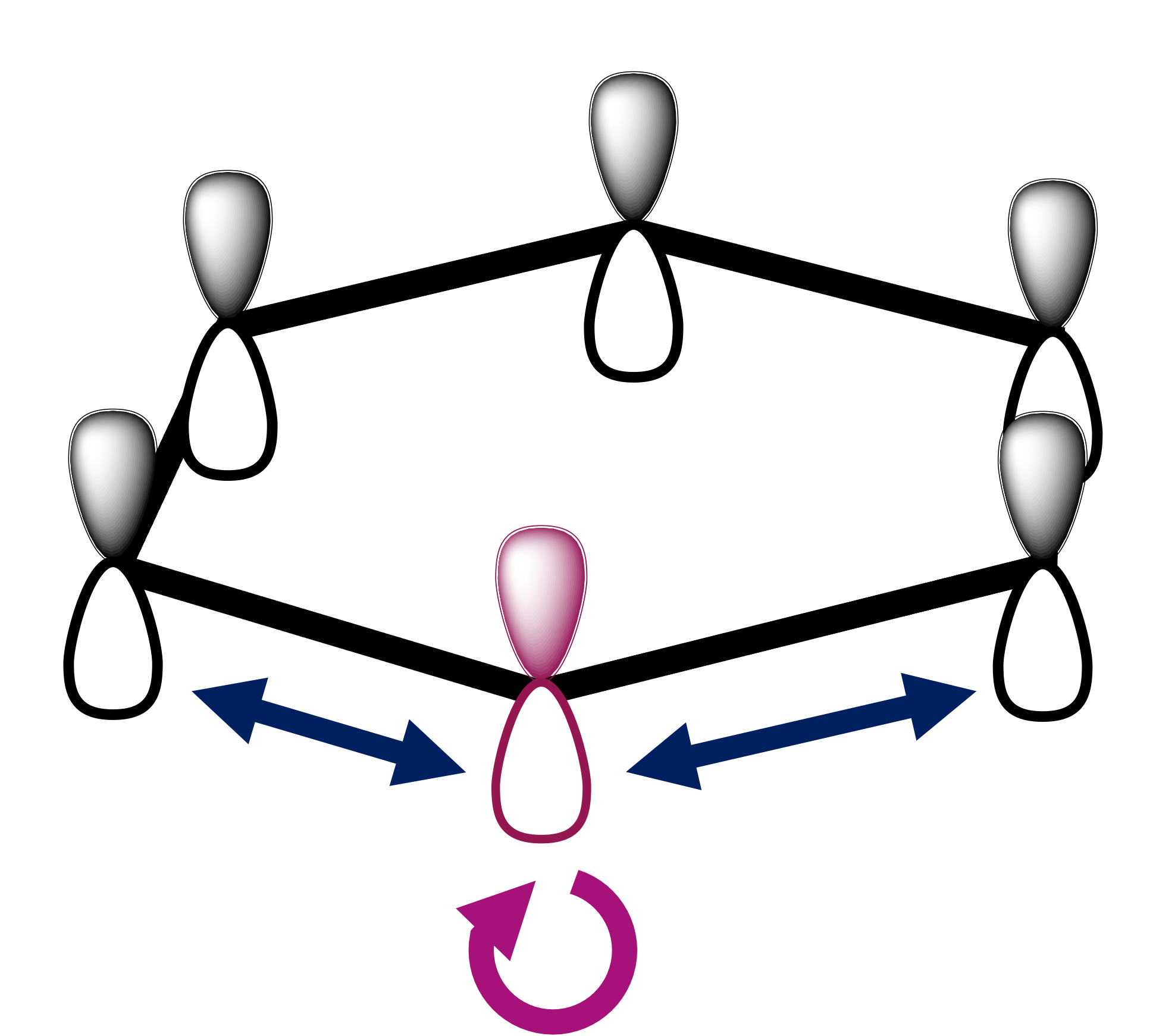}
         \caption{Hubbard}
         \label{fig::Hubbard}
     \end{subfigure}
     \hfill
     \begin{subfigure}[b]{0.23\textwidth}
         \centering
         \includegraphics[width=\textwidth]{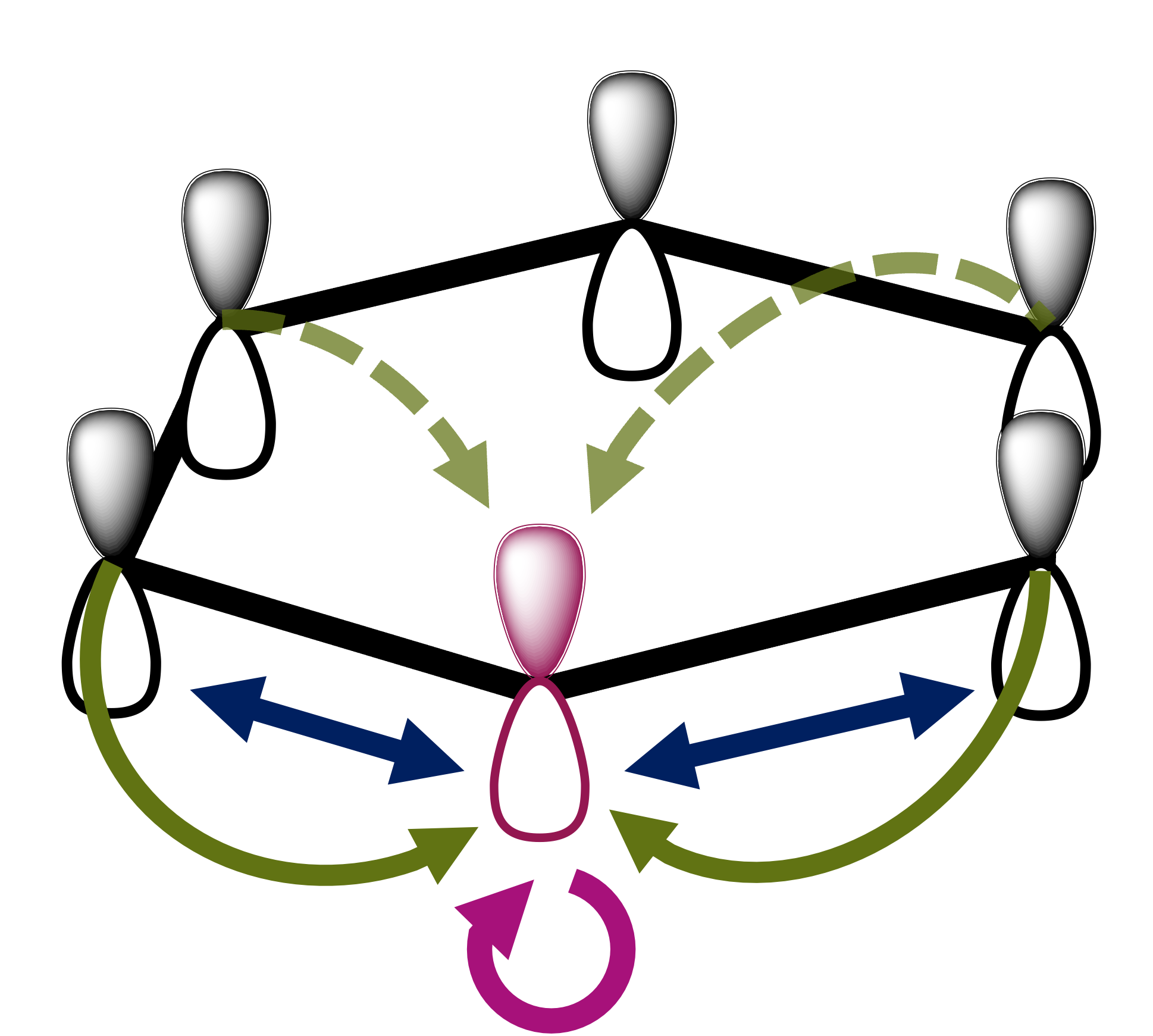}
         \caption{extended Hubbard}
         \label{fig::extHubbard}
     \end{subfigure}
     \hfill
     \begin{subfigure}[b]{0.23\textwidth}
         \centering
         \includegraphics[width=\textwidth]{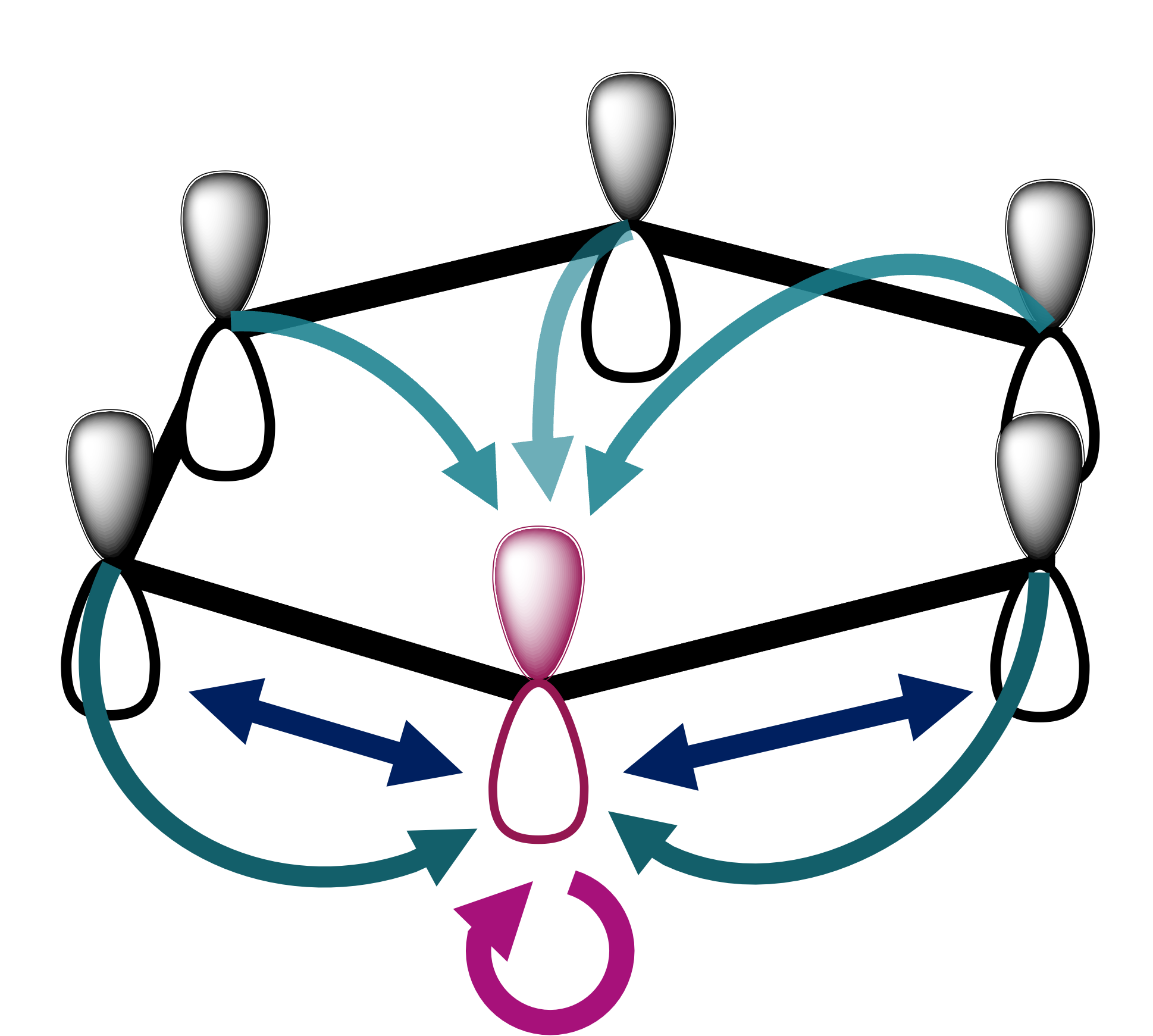}
         \caption{PPP}
         \label{fig::PPP}
     \end{subfigure}
     \caption{Illustration of the different model Hamiltonians on benzene. For each model, all interactions with respect to a specific \mbox{$\pi$-orbital} (highlighted in purple) are depicted. In the Hückel Hamiltonian, only hopping terms (dark blue) are included. In the Hubbard model, an on-site interaction term (purple) is added for each site. By also including parametrized interactions between nearest neighbours (solid green) and optionally also next-nearest neighbours (dashed green), one obtains the extended Hubbard Hamiltonian. In the PPP model, interaction terms between all sites are included (teal), and the interaction strength is scaled based on the inter-atomic distance between the sites.}
     \label{fig::ModelHamiltonians}
\end{figure*}

We start from the Hückel Hamiltonian, as it forms the base for many more advanced model Hamiltonians.
We define the model in second quantized notation as 
\begin{equation}
\begin{split}
    H_\textnormal{Hück}(\epsilon,t) &=  \sum_{i}\epsilon_i~\hat{n}_i-\sum^{'}_{i,j,\sigma} t_{ij}~(\hat{a}^{\dagger}_{i\sigma}\hat{a}_{j\sigma}+\hat{a}^{\dagger}_{j\sigma}\hat{a}_{i\sigma})\\ 
    &=  H_0(\{\epsilon,t\}^\textnormal{Hück}) \label{eq:Hueck}
    \end{split}
\end{equation}
where $\hat{a}^{\dagger}_{i\sigma}$ ($\hat{a}_{i\sigma}$) creates (annihilates) an electron with spin $\sigma$ in the $p_z$-AO located on atom $i$ and $\hat{n}_i=\sum_\sigma \hat{a}^\dagger_{i\sigma}\hat{a}_{i\sigma}$ counts the total number of electrons on atom $i$.
The onsite orbital energy $\epsilon_i$, also known as the core resonance integral $\alpha$, is often discarded as it only accounts for a constant energy shift when all atoms are equivalent.
This is true for idealized geometries, (\textit{i.e.} equal bond lengths) that do not contain any heteroatoms such as nitrogen or oxygen \cite{kutzelnigg_einfuhrung_2002}. 
The second term, $t_{ij}^{\textnormal{Hück}}$ is the kinetic energy or hopping integral, also known as the resonance integral $\beta$ in the chemistry literature.
The primed sum only runs over connected atoms $i$ and $j$ (i.e. directly bonded).
An illustration of this interaction for benzene is shown in Fig.~\ref{fig::Hueckel}.
In solid-state physics, equation~\eqref{eq:Hueck} is also known as the tight-binding model.

The Hückel Hamiltonian is a one-particle Hamiltonian that can also comprise the non-interacting part $H_0$ of a more general interacting Hamiltonian, with a specific parameter set $\{\epsilon,t\}^\textrm{Hück}$.
This Hamiltonian does not explicitly include electron-electron interactions, neither between electrons on the same atom $i$, with different spin $\sigma$, nor between electrons on different atoms.
This omission limits the model's ability to accurately describe polar bonding \cite{kutzelnigg_einfuhrung_2002} and to predict the correct singlet-triplet energy splitting for a given electron configuration \cite{parr_quantum_1972}, both of which are chemically significant.

In polymers, these limitations are also evident: the Hückel Hamiltonian fails to reproduce experimentally observed features such as negative spin densities in linear polyenes \cite{ramasesha_correlated_1984, ma_density_2022}, topological solitons in odd polyenes \cite{ramasesha_correlated_1984}, and nonlinear optical properties in conjugated polyenes \cite{ramasesha_second_1990}.
Nevertheless, the Hückel Hamiltonian has been successful in providing a realistic description of charge mobility in conjugated polymer systems \cite{berencei_realistic_2019}, after the inclusion of electron-phonon coupling terms, leading to the Holstein model.
Overall, however, it has to be noted that the neglect of explicit electron-electron interaction in the Hückel model is often too drastic an approximation to answer chemically relevant questions.

\subsubsection{Hubbard and extended Hubbard Hamiltonian}
\label{subsubsec::Hubbard}
To overcome the shortcomings of the Hückel model, one can try to improve this description through the inclusion of some electron-electron interaction.
Starting from the non-interacting Hamiltonian $H_0$ and adding the electron-electron interaction term $U$, defined as the interaction between two electrons on the same atom, will yield the Hubbard Hamiltonian (see also Fig.~\ref{fig::Hubbard}):
\begin{equation}
    H_\textnormal{Hub}(\{\epsilon,t,U\}^\textnormal{Hub})=H_0(\epsilon,t)+\frac{U}{2}\sum_{i}\hat{n}_{i}(\hat{n}_{i}-1), \label{eq:Hub}
\end{equation}
where we have denoted the model parameters for $H_\textnormal{Hub}$ as a unique set $\{\epsilon, t,U\}^\textnormal{Hub}$ because parameters should not generally be transferred from one model Hamiltonian to another \cite{chiappe_can_2015,skotheim_electroresponsive_1988}.

The Hubbard model has been widely applied in solid-state physics, but has seen more limited use in chemical contexts \cite{kutzelnigg_what_2007}.
The most common application of Hubbard-type models in chemistry is the DFT+U method, a density functional theory (DFT) approach, where the $U$ parameter is included in the functional form for strongly correlated systems. 
In conjugated $\pi$-electron systems, long-range electron-electron interactions often play a crucial role\cite{gundra_pariserparrpople_2013}, which are not described by the Hubbard model. This has qualitatively significant consequences, for instance, the standard Hubbard model cannot describe a bound exciton in polymers (a bound state of an electron and a hole) due to the restriction to local, same-atom electron-electron interactions characterized by $U$ \cite{ma_density_2022}.

To describe bound excitons with the Hubbard model, some form of inter-atomic electronic interaction is necessary and one usually defines an extended Hubbard Hamiltonian (Figure~\ref{fig::extHubbard})
\begin{equation}
\begin{split}
    H_\textnormal{exHub}(&\{\epsilon,t,U,V\}^\textnormal{Hub}) = \\ 
    &H_0(\epsilon,t)~+~\frac{U}{2}\sum_{i}\hat{n}_{i}(\hat{n}_{i}-1)+\frac{1}{2}\sum^{'}_{i \neq j}V_{ij}(n_i-z_i)(n_j-z_j), \label{eq:exHub}
    \end{split}
\end{equation}
where the second sum usually runs over nearest and optionally next-nearest neighbour atoms $i$ and $j$.
The effective charge $z$ of the atomic core $i$, evaluated when $\pi$-electrons are removed (so $z_i=1$ for carbon atoms) is often not considered for infinite systems, but must be included for finite systems\cite{chiappe_can_2015}.
It has been shown, however, that even the extended Hubbard model does not predict bound excitons in conjugated polymers\cite{shuai_binding_1997,shuai_exciton_1998} and that long-range electron-electron interactions, as included in the PPP Hamiltonian, are essential for their description\cite{ma_density_2022}. 

\subsubsection{PPP Hamiltonian}
\label{subsubsec::PPP}

Finally, the PPP Hamiltonian (Figure~\ref{fig::PPP}) with the inclusion of long-range Coulomb interaction between all atoms is given as
\begin{equation}
    \begin{split}
        H_{PPP}(& \{\epsilon,t,U,V\}^\textnormal{PPP})=\\
        &H_0(\epsilon,t)+\frac{U}{2}\sum_{i}\hat{n}_{i}(\hat{n}_{i}-1)+\frac{1}{2}\sum_{i \neq j}V_{ij}(\hat{n}_i-z_i)(\hat{n}_j-z_j),\label{eq:PPP}
    \end{split}
\end{equation}
where the only difference from the extended Hubbard model in equation~\eqref{eq:exHub}, apart from the different parameter set, is that restriction to nearest or next-nearest neighbour atoms in the third term is lifted.
The inter-atomic electron-electron interaction in the PPP Hamiltonian is often parametrized either through the Ohno potential\cite{ohno_remarks_1964}
\begin{equation}
    \label{eq:Ohno}
    V_{ij} = \frac{U}{\sqrt{1+(U\epsilon_r r_{ij}/14.397)^2}},
\end{equation}
or the Mataga-Nishimoto potential \cite{mataga_electronic_1957}
\begin{equation}
    \label{eq:MatagaNishimoto}
        V_{ij} = \frac{U}{1+U\epsilon_r r_{ij}/14.397},
\end{equation}
where both potentials are interpolations between a Coulomb potential at long distances (given in Å) and the U parameter (given in eV) at short distances $r_{ij}$ and therefore do not introduce another parameter, as one might assume from equation~\eqref{eq:PPP}.
The relative permittivity $\epsilon_r$ is commonly set to unity.

When comparing the different models, we can say that in the weak coupling limit, where $U/t\ll 1$, all model Hamiltonians become Hückel-like.
A detailed comparison between the PPP and Hubbard Hamiltonians for finite graphene and polycyclic aromatic hydrocarbons (PAH) found that the PPP Hamiltonian succeeds in accounting for long-range interactions. These interactions effectively screen ionic charges, whereas the standard and extended Hubbard Hamiltonian can fail to capture this behaviour\cite{chiappe_can_2015}.
In conclusion, the PPP model has been found to be generally better suited for describing the electronic properties of PAHs and related conjugated systems than the Hubbard models or even simpler Hückel Hamiltonians\cite{chiappe_can_2015}.

\subsection{Validation}
\label{subsec::Validation}

The PPP Hamiltonian is semi-empirical in its nature and in theory less general than a full \textit{ab\,initio} treatment.
In practice, the much more involved and computationally expensive \textit{ab\,initio} methods often struggled to match the agreement with experiment as obtained through simpler and faster PPP model calculations\cite{brandow_formal_1979}.
Of course, one could argue that this is unsurprising, given that semi-empirical PPP parameters represent a fit to experimental data.
However, this argument falls short because the PPP parameters for a specific small molecular system (e.g. ethene or benzene) have been shown to transfer well to other systems far outside the original parametrization regime.
Explaining this success of the PPP model from a theoretical standpoint has posed a great challenge. In general, the theoretical justifications for the various approximations have been given much later, well after their ad hoc introduction and demonstrated success.
One of the reasons for this delay, has been that detailed \textit{ab\,initio} calculations were computationally demanding and only once they became feasible, was it possible to mimic these approximations within the \textit{ab\,initio} framework.
Here we present the most consequential approximations invoked in the PPP model, namely $\pi$-electron treatment, zero differential overlap, and semi-empirical parameters, and discuss their theoretical justification and limitations. 

\subsubsection{\texorpdfstring{$\pi$}{π}-electron systems}
We have already discussed a distinction between $\sigma$- and $\pi$-MOs in the case of planar conjugated molecules for HMO theory and in the PPP model.
In the $\pi$-electron approximation electrons in $\sigma$-MOs are not treated explicitly but their effective interactions with the $\pi$-electrons in the corresponding MOs enter through the model parametrization.
The $\pi$-electron approximation therefore results in a significant reduction from an all-electron space to just the valence $\pi$-electron subspace.
Both the Hückel and the PPP model thus align naturally with the chemical intuition of organic chemists, where predominantly the $\pi$-bonds are considered to describe chemical reactivity.

A theoretical validation for reducing the full electronic Hamiltonian to a valence $\pi$-electron Hamiltonian, is given by the effective valence shell Hamiltonian $\mathscr{H}^v$ approach \cite{brandow_formal_1979,segal_semiempirical_1977-1,martin_ab_1993,martin_ab_1994,martin_ab_1994-2,martin_ab_1994-1,martin_ab_1995}.
Here the complete molecular electronic Hamiltonian is cast into a formally exact effective Hamiltonian which acts solely within the valence space \cite{martin_ab_1996}.
This approach can mimic the approximations made for the PPP Hamiltonian, such as the $\pi$-electron approximation, through an \textit{ab\,initio} Hamiltonian and evaluate their merit.
Within this approach, it has been validated that, although not explicitly included in the PPP Hamiltonian, $\sigma$- and $\pi$-orbital relaxation, $\sigma-\sigma$, $\pi-\pi$ and $\sigma-\pi$-correlations are included through adjustment of parameters when accounting for correlation effects \cite{martin_ab_1994-1}.
Furthermore, the $\pi$-electron approximation within the \textit{ab initio} effective Hamiltonian $\mathscr{H}^v$ is exact and generally also holds for the semi-empirical Hamiltonian.
This is true as long as complicated effective n-body interaction terms with $\textrm{n}\ge3$ can be neglected for the \textit{ab\,initio} Hamiltonian, or alternatively included in an averaged fashion in the PPP Hamiltonian\cite{brandow_formal_1979,andre_linked-cluster_1978,martin_ab_1994-1}.

\subsubsection{Zero Differential Overlap}
\label{subsubsec::ZDO}
The zero differential overlap (ZDO) approximation assumes orthonormal AOs and results in a very reduced set of integrals that have to be considered. 
Within the ZDO approximation there are no two-body two-centre resonances, hybrid and exchange integrals, or three- and four-centre integrals \cite{martin_ab_1994-1}.
The kinetic energy integrals are usually assumed to include only nearest-neighbour interactions.
This very reduced description is commonly explained by recasting the basis over symmetrically orthogonalized Löwdin orbitals, which only truly holds when correlation contributions are included \cite{martin_ab_1995}.
Effective valence Hamiltonian $\mathscr{H}^v$ calculations on benzene and cyclobutadiene show that in the case where correlation contributions are included correctly most two-electron resonance, hybrid, and exchange matrix integrals become very small.
Furthermore, almost all three- and four-centre two-electron integrals can be considered negligible, corroborating the ZDO approximation for the PPP Hamiltonian\cite{martin_ab_1995}.
For the kinetic energy integrals, it is found that non-nearest neighbour integrals can be non-negligible\cite{martin_ab_1995,martin_ab_1996}.

\subsubsection{Transferability of semi-empirical parameters}
\label{subsubsection::SemiEmpirical}
The transferability of parameters between different systems is a core objective for a semi-empirical approach, and often implicitly assumed.
As already discussed in section~\ref{subsec::SemiEmpirical}, semi-empirical calculations can fail dramatically when used outside the valid parametrization range.
It is therefore relevant to theoretically understand how transferability can be maximized.
An important validation of the semi-empirical parameters and their transferability for the PPP Hamiltonian has been provided by Freed and coworkers in their extensive work on the effective valence Hamiltonian $\mathscr{H}^v$\cite{segal_semiempirical_1977-1,martin_ab_1993,martin_ab_1994,martin_ab_1994-2,martin_ab_1994-1,martin_ab_1995}.
They showed that for $\mathscr{H}^v$ calculations of ethylene, trans-butadiene, cyclobutadiene, hexatriene and benzene, remarkable transferability of correlated integrals could be obtained \cite{martin_ab_1994-1,martin_ab_1995}.
However, this transferability critically depends on the correct inclusion of correlation interaction.
The importance of correlation for the transferability of the parameters had already been noted earlier, specifically for the parametrization of electron-electron interaction\cite{schulten_correlation_1976}.

From a computational perspective, the need to include considerable electron correlation for the approximations underlying the PPP Hamiltonian to hold and thereby reach accurate results poses a significant challenge.
A full configuration interaction (FCI, also called exact diagonalization in physics) calculation will quickly become unfeasible due to its exponential scaling, even for a reduced Hilbert space such as with the PPP Hamiltonian.
Even a more restricted inclusion of correlation, where excitations in the configurations of a reference wave function are considered to some order (single, double, triple excitation etc.) will become intractable for larger systems due to their generally high polynomial scaling. 
On the other hand, once a complete description of electron correlation within the model can be reached, a ``model exact study'', the true validity of the model and its parameters can be assessed, and discrepancies from experimental data can be used to analyse what is missing in the model description.
For early examples of this model exact approach, see for instance Refs.~\citenum{visscher_exact_1970, ducasse_correlated_1982, ramasesha_correlated_1984, soos_valence_1989, ramasesha_second_1990, ramasesha_optical_1991, albert_linear_1992, ramasesha_exact_1993}

\section{PPP then}
\label{section::PPPThen}
\begin{figure}[htb]
    \centering
    \includegraphics[width=0.28\textwidth]{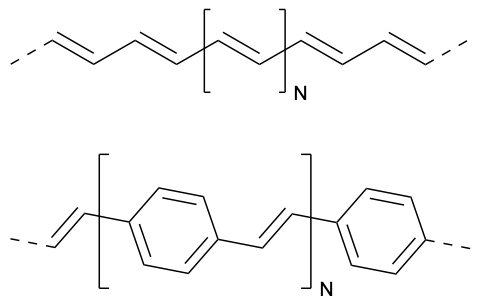}
    \caption{\textit{trans}-polyacetylene (top) and poly(\textit{para}-phenylene-vinylene) (bottom)}
    \label{fig::Polymers}
\end{figure}

In this section, we will illustrate why and how the PPP Hamiltonian has remained relevant over so many years, even after its original intended use in small conjugated molecules was subsequently taken over by more general full \textit{ab\,initio} methods.
We focus on conjugated polymers to showcase this development.
Conjugated polymers are quasi-one-dimensional systems where the PPP Hamiltonian has been widely applied with great success.
Two exemplary polymers,\textit{trans}-polyacetylene and poly(\textit{para}-phenylene-vinylene) are shown in Fig.~\ref{fig::Polymers}.

Increased theoretical interest in polymers arose after the successful synthesis of thin polyacetylene films \cite{ito_simultaneous_1974} and following the discovery of conductance for doped polyacetylene \cite{shirakawa_synthesis_1977,chiang_electrical_1977}, which was awarded the Nobel Prize in Chemistry in 2000.
These findings opened up a large field of technical applications such as plastic field-effect transistors, electromagnetic shielding, nonlinear optical devices, photovoltaic devices and light-emitting devices \cite{barford_electronic_2013}, in particular, following the discovery of electroluminescence in phenyl-based polymers, such as poly(\textit{para}-phenylene-vinylene)\cite{burroughes_light-emitting_1990}.

The theoretical understanding of such polymers can be challenging due to the extended size of these systems.
Furthermore, electron-electron interactions are only weakly screened (due to the low dimensionality of the system) and therefore electron correlation has to be considered\cite{barford_electronic_2013}.
The interaction strength is commonly characterized as weak in phenyl-based polymers such as poly(\textit{para}-phenylene-vinylene) and as intermediate, in between the weak and strong electron-electron interaction limits, for polyenes such as \textit{trans}-polyacetylene\cite{barford_electronic_2013}.

While an \textit{ab\,initio} treatment with correlated methods would be the preferred choice in case of strong electron-electron interaction, this becomes very challenging due to the large system sizes one has to reach to extrapolate to the infinite system size limit \cite{ramasesha_optical_1991}.
The simpler independent particle description with single-reference \textit{ab\,initio} methods such as RHF, UHF or DFT will not be sufficient for the intermediate-coupling regime.
It has therefore been advantageous to revert to different model Hamiltonians, where we will discuss here specifically results with the PPP Hamiltonian.
Our primary focus is the model exact approach because it unambiguously showcases the strengths and limitations of a chosen model separated from any potential approximations invoked in the method used to solve it\cite{cardona_overview_1992}.
Finally, we will also discuss an approximate FCI-type solver, the density matrix renormalization group (DMRG) method, and how it benefitted uniquely from the PPP Hamiltonian in describing polymeric systems.

\subsection{Model exact studies}
\label{subsec::ModelExactStudies}
We have already seen in section~\ref{subsec::Validation} that electron correlation has to be considered for the fundamental approximations of the PPP Hamiltonian to hold.
A valid question one can ask then is how much correlation needs to be included to obtain reliable results? An early comparison between various levels of perturbation theory and FCI calculations on the electronic structure of benzene found correlation to be important, and the convergence in the perturbation series towards the FCI result to be slow in terms of excitations \cite{visscher_exact_1970}. 
Additional evidence for the importance of electron correlation was established for linear polyenes through experimental observation \cite{hudson_low-lying_1972} and subsequent theoretical confirmation \cite{schulten_origin_1972,hudson_linear_1982}.

Based on these findings, another key point was raised regarding the influence of the interaction strength on the necessary level of electron correlation \cite{schulten_correlation_1976}: 
Depending on the interaction strength in a system, either of the two respective reference basis representations (MO and valence bond (VB) basis) can become inadequate to resolve polyene spectra, unless solved for FCI, where both approaches become equivalent\cite{hudson_linear_1982}.
As already discussed earlier, polyenes usually fall in the intermediate interaction regime \cite{barford_electronic_2013,ma_density_2022} and model exact treatment with FCI can therefore give reassurance that interactions are adequately described.

\subsubsection{Size extensivity}
\label{subsubsec::SizeExtensivity}
The model exact approach for polyenes and polymers is not only important for the correct characterisation and ordering of electronic states, but additionally for its inherent size extensivity. Size extensivity is achieved when the energy of the system scales correctly with the increasing number of repeated units.
Due to the large system sizes of conjugated polymers, any correlated method would need to treat smaller oligomers first and extrapolate from these results to the polymers\cite{ma_density_2022}.
This can however only be done faithfully with correlated methods that are size extensive. 

While FCI is size extensive, this property is lost if the CI expansion is truncated, as is the case with configuration interaction singles and doubles (CISD) \cite{ramasesha_optical_1991}.
There are, however, other widely used correlated methods such as coupled cluster (CC) that are size extensive and have been employed with the PPP Hamiltonian to calculate infinite-chain properties from cyclic polyenes by extrapolating from smaller chain lengths \cite{paldus_coupled-cluster_1984}. 
For these systems, single-reference CC calculations break down, when correlation effects become sufficiently strong in larger polyenes\cite{podeszwa_electronic_2002}.
This is especially true when a mean-field solution is taken as the reference state. Extensions to the CC approach with a valence bond reference \cite{paldus_pppvb_1991,paldus_valence_1994,planelles_valence_1994-1,planelles_valence_1994,paldus_valence_1999}, on the other hand, have been shown to work well, also in the highly correlated limit.

\subsubsection{Validation of parameters}
\label{subsubsec::FocusParameters}
The model exact approach describes the entire energy spectrum exactly within the chosen basis.
For the PPP Hamiltonian, this amounts to the $\pi$-electron subspace and all $\pi-\pi^*$ excitations, which can be compared with experimental results.
The model exact approach therefore shifts focus away from the amount of correlation included, to the validity of the parameters and the model itself \cite{ramasesha_exact_1993}.
Only then it is clear if the approximations made are appropriate.
Various FCI calculations with the PPP Hamiltonian have been performed for a diverse set of conjugated $\pi$-electron systems \cite{ramasesha_correlated_1984,ramasesha_optical_1991,ramasesha_exact_1993} showing that the standard parameters can qualitatively reproduce the lowest excited states.

The original parametrization predated extensive CI calculations \cite{soos_exact_1983} and was based on experimental data and either SCF or limited CI calculations \cite{ramasesha_exact_1993}.
The standard parameters were considered to demonstrate robustness and transferability \cite{ramasesha_exact_1993} because they hold up well once electron correlation is accounted for. The proper inclusion of electron correlation is thus preferable over reparameterisation of the model as previously undertaken for mean-field approaches\cite{skotheim_electroresponsive_1988}.

Later, screened parameters were proposed\cite{chandross_coulomb_1997} that better describe the high-energy excited states of phenyl-based polymers \cite{shukla_theory_2003}.
The screening was originally suggested as an environmental effect in the condensed phase, which reduces the effective charge of $\pi$-electrons over larger distances.
Further model exact calculations on phenyl-systems \cite{castleton_screening_2002} concluded, however, that the screening parameters had to be included even in the gas phase.
The screening effect was thus attributed to the screening effect of the $\sigma$-electrons rather than external environment effects\cite{castleton_screening_2002}.

\subsubsection{Validation of model}
\label{subsubsec::FocusModel}
The model exact approach not only allows validation of the chosen parameters, but also of the model itself.
With highly accurate calculations such as FCI, the ambition often became to reach a sub-0.1eV accuracy in the theoretical assignment of experimental spectra\cite{ramasesha_exact_1993}.
This is a challenging demand because many small effects can play a role in experiments and might not easily be included in the model.

As a concrete example, we mention symmetry-adapted FCI calculations of the anthracene molecule, essentially three fused benzene molecules along one axis, where a range of effects \cite{ramasesha_exact_1993} prohibited the PPP model description from reaching sub-0.1eV accuracy.
The first of the effects is symmetry-breaking, which is not captured by the symmetry-adapted basis that is utilized.
Another effect is vibronic coupling or nonadiabatic coupling, which is neglected in the PPP model due to the underlying Born-Oppenheimer (BO) approximation.
In the BO approximation, the motion of atomic nuclei are assumed to be decoupled from the motion of electrons and only affect the electrons parametrically.
This is usually an excellent approximation, but can fail when electronic states become degenerate or nuclear motion relative to the electronic timescale cannot be neglected \cite{barford_breakdown_2002}.

An additional source of potential discrepancy between model exact studies and experimental data can be the relaxation of the molecular structure in the excited state from the ground state geometry.
While this effect was considered in the anthracene study \cite{ramasesha_exact_1993}, the authors caution that for accurate relaxation energies the difference between solid-state experimental data and gas-phase molecular calculations has to be considered as a solid-state shift in energy of up to $0.2-0.3$eV.  

While every single effect described here can potentially be addressed and corrected for in a model, solving the augmented model computationally becomes harder and harder.
The hope then is to understand a model and its limitations confidently and exactly for smaller systems, so that the approximations can be made with reassurance in bigger systems.
Researchers can differ significantly in their perception of such models, which also influences how they are employed and benchmarked.  Some believe that exploring the model itself holds value, and in this context the model exact approach is particularly interesting. There are also others who view the model as simply a tool for understanding the experimental reality, thus being more interested in the reproduction of experimental results. In this case, the model exact approach can provide the benchmark data to support the use of more approximate methods to focus on questions of experimental interest. 

\subsection{Beyond model exact studies}
\subsubsection{Benchmark for other correlated methods}
\label{subsubsec::Benchmark}
One of the most important uses for a model exact approach is the validation of more approximate methods.
The PPP Hamiltonian in the model exact approach has been used extensively to benchmark the accuracy of various methods, and has been used for the development of new methods as well\cite{chuiko_modelhamiltonian_2024}.
We will discuss the significant role that the PPP Hamiltonian played in the introduction of the density matrix renormalization group (DMRG) technique into the chemistry community in the next section. In this section, we focus on the extensive work of Paldus and coworkers, who used the PPP Hamiltonian \cite{cizek_cluster_1969,paldus_correlation_1974,paldus_cluster_1982,takahashi_perturbation_1983,paldus_degeneracy_1984,paldus_coupled-cluster_1984,takahashi_coupled-cluster_1985,piecuch_coupled_1990,piecuch_coupled-cluster_1990,piecuch_coupled-cluster_1990-1,piecuch_solution_1991,paldus_electron_1992,piecuch_behavior_1992,paldus_valence_1994,planelles_valence_1994-1,planelles_valence_1994,paldus_valence_1999} for various coupled cluster approaches.

In an extensive body of work, Paldus and coworkers showed that the CCSD approach breaks down for annulenes (strongly correlated  cyclic polyenes) and conceived an approximate coupled-pair method (ACPQ) that could deliver close to model exact results for these systems. \cite{podeszwa_electronic_2002}.
Later, the valence bond coupled cluster approach with singly and doubly excited cluster amplitudes (VB-CCSD) was developed and tested by the same group for the PPP Hamiltonian on $\pi$-electron systems, especially for strongly correlated systems.
In a severely resource-constrained compute environment, the PPP Hamiltonian offered a way to check exact solutions and furthermore allowed the exploration of the entire range of correlation effects, simply by tuning the hopping term/resonance integral \cite{planelles_valence_1994}. Together, these features meant that the PPP Hamiltonian played a significant role in the methodological development of the coupled cluster approach\cite{paldus_beginnings_2005}.

\subsubsection{Beyond the reach of FCI with the density matrix renormalization group}
\label{subsec::DMRG}
One of the greatest challenges with the model exact approach is that the computational effort for an FCI calculation will scale exponentially with the increasing system size.
Different strategies have been employed to reduce the Hilbert space size of the problem and push the onset of computational infeasibility further out.
Such techniques include employing a reduced model Hamiltonian, such as the PPP Hamiltonian, or exploiting the symmetries of the Hamiltonian, leading to a block-diagonalizable Hamiltonian.
Eventually, however, exponential scaling is inevitable.

For conjugated polyenes and more complex polymer systems, we have seen throughout this chapter that accurate treatment of correlation is essential.
Furthermore, there was considerable interest in reaching much larger system sizes than what could be afforded by an FCI treatment.
Consequently, the adaptation of the density matrix renormalization group method (DMRG) for chemistry in the context of model Hamiltonians in the 1990s marked a great opportunity to tackle conjugated systems beyond what was previously possible\cite{ma_density_2022}.

There are several reasons why  DMRG using the PPP Hamiltonian is particularly useful for the treatment of conjugated polymers.
DMRG is size extensive for one dimension \cite{weatherly_embedded_2025}, variational, and therefore in theory systematically improvable until FCI accuracy.
Furthermore, the computational scaling is reduced due to the sparsity of the PPP Hamiltonian\cite{bursill_symmetry-adapted_2009}.
An additional advantage for DMRG is the particle-hole symmetry of the PPP Hamiltonian, which can be exploited together with spatial and spin-flip symmetries to target high-lying excited states \cite{bursill_symmetry-adapted_2009}.
For these reasons, unprecedented lengths of polymer chains containing 100 carbon atoms could be accurately investigated using DMRG \cite{ma_density_2022}.
This answered long-standing questions in excited state ordering, exciton binding energies, and solved problems in linear and nonlinear spectroscopy of conjugated polymer systems.
It also conclusively confirmed the usefulness of DMRG for chemistry, which was then later also extended to \textit{ab initio} Hamiltonians. \cite{ma_density_2022} 

\section{PPP today}
\begin{figure}[htb]
    \centering
    \includegraphics[width=0.8\textwidth]{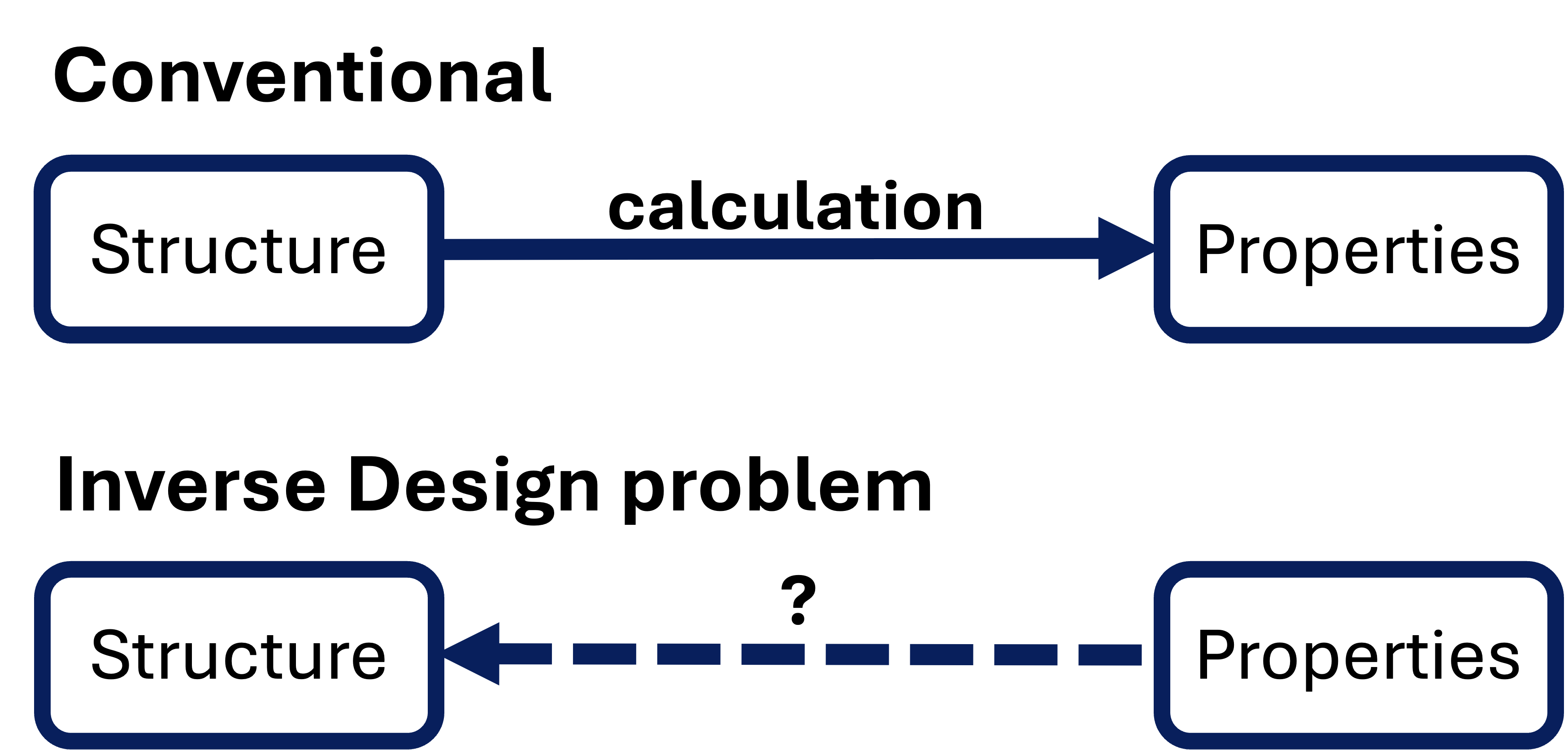}
    \caption{Inverse design problem, adapted from Ref.~\citenum{green_inverse_2022}}
    \label{fig:InverseDesignProblem}
\end{figure}

In the previous section, we have shown how the relatively simple PPP model has helped to gain theoretical understanding of experimentally relevant electronic and optical properties in conjugated polymers.
This process can be viewed as gaining understanding from a given structure (molecule, polymer, etc.) through electronic structure calculations and is shown in the top of Fig.~\ref{fig:InverseDesignProblem}.
The results of these calculations are properties of the system of interest, such as the electronic structure, optical spectra or similar.
These properties can be compared with experimental data and may be refined through more accurate calculations or a better model.
While this process has been successful in gaining insight, it can also be relatively slow because it involves many calculations.

There is also no guarantee that the calculated properties are useful or even insightful for experimental chemists searching for new systems of interest.
These molecular engineering problems could, for example, include the search for a better catalyst or a more efficient photovoltaic materials.
Ideally, calculations could also deliver insight for the opposite problem, as indicated in Fig.~\ref{fig:InverseDesignProblem}: from some defined properties to a new and improved structure.
This is known as the inverse design problem \cite{sanchez-lengeling_inverse_2018,cartwright_artificial_2020,green_inverse_2022} and has been discussed extensively in the context of high-throughput screening. 

One of the challenges inherent in the inverse design problem is the vastness of chemical space, by some estimates consisting of $10^{60}$ molecules \cite{cartwright_artificial_2020}.
Often, researchers do not feel completely in the dark and believe that chemical intuition can reduce this space considerably to some of the most relevant regions.
Even if this is true, the question is how human chemical intuition can be translated into actionable instructions for a computer.
One strategy that has been employed is the use of design rules to significantly reduce the chemical space before the computationally intensive search using virtual screening, machine learning, and genetic algorithms. \cite{green_inverse_2022}.
These design rules, or simply chemical intuition, have a long history in organic chemistry. One of the early examples is Hückel's rule, which is derived from HMO theory and predicts compounds with $(4n+2)$ $\pi$-electrons to be aromatic \cite{huckel_quantentheoretische_1932}.

The PPP Hamiltonian has been discussed in this context as a tool to derive a cheap scoring function\cite{jorner_ultrafast_2024} and has been used as a tool to formulate and verify design rules\cite{hele_anticipating_2019,dubbini_turning_2024,green_exroppp_2024,valentim_simple_2020,green_inverse_2022,savi_organic_2025}.
The PPP Hamiltonian has been employed in the construction of design rules, especially for its ability to treat larger system sizes than otherwise would be possible with \textit{ab\,initio} multi-configurational methods\cite{hele_anticipating_2019}.
Furthermore, the particle-hole symmetry of the PPP Hamiltonian has proved useful in the determination of the energetic ordering and form of molecular orbitals in radical emitter systems, but particle-hole symmetry is broken by chemical substitution \cite{dubbini_turning_2024}.
Proposals include specifically tailored design rules for systems such as acenes \cite{hele_anticipating_2019}, radical-based organic light-emitting diodes (OLEDs) \cite{green_exroppp_2024}, magnetic molecules \cite{valentim_simple_2020} and a more generalized framework description of design rules for the PPP Hamiltonian\cite{green_inverse_2022}.

We focus our discussion in the following sections on two relevant application fields, singlet fission and the search for molecules with an inverted singlet-triplet energy gap (InveST).
These constitute two very challenging inverse design problems, and we will describe the challenges in modelling and how the PPP Hamiltonian specifically has been used to gain insight and formulate design rules.

\subsection{Singlet fission}
\label{subsection::singletfission}
\begin{figure}[htb]
    \centering

    \begin{subfigure}[b]{0.4\textwidth}
         \centering
         \includegraphics[width=\textwidth]{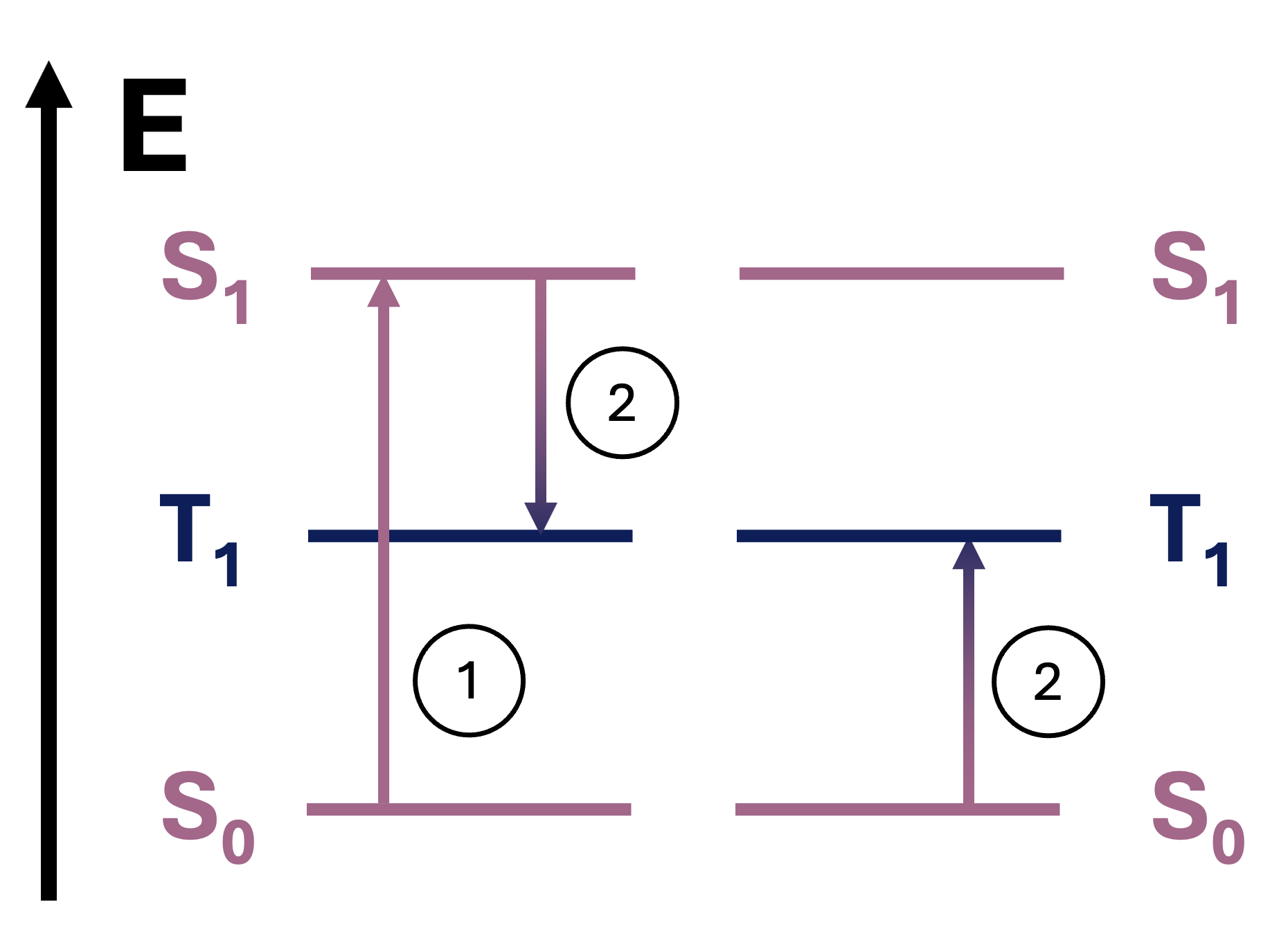}
         \caption{Singlet Fission}
         \label{fig::SF}
     \end{subfigure}
     \hfill
     \begin{subfigure}[b]{0.44\textwidth}
     \centering
     \includegraphics[width=\textwidth]{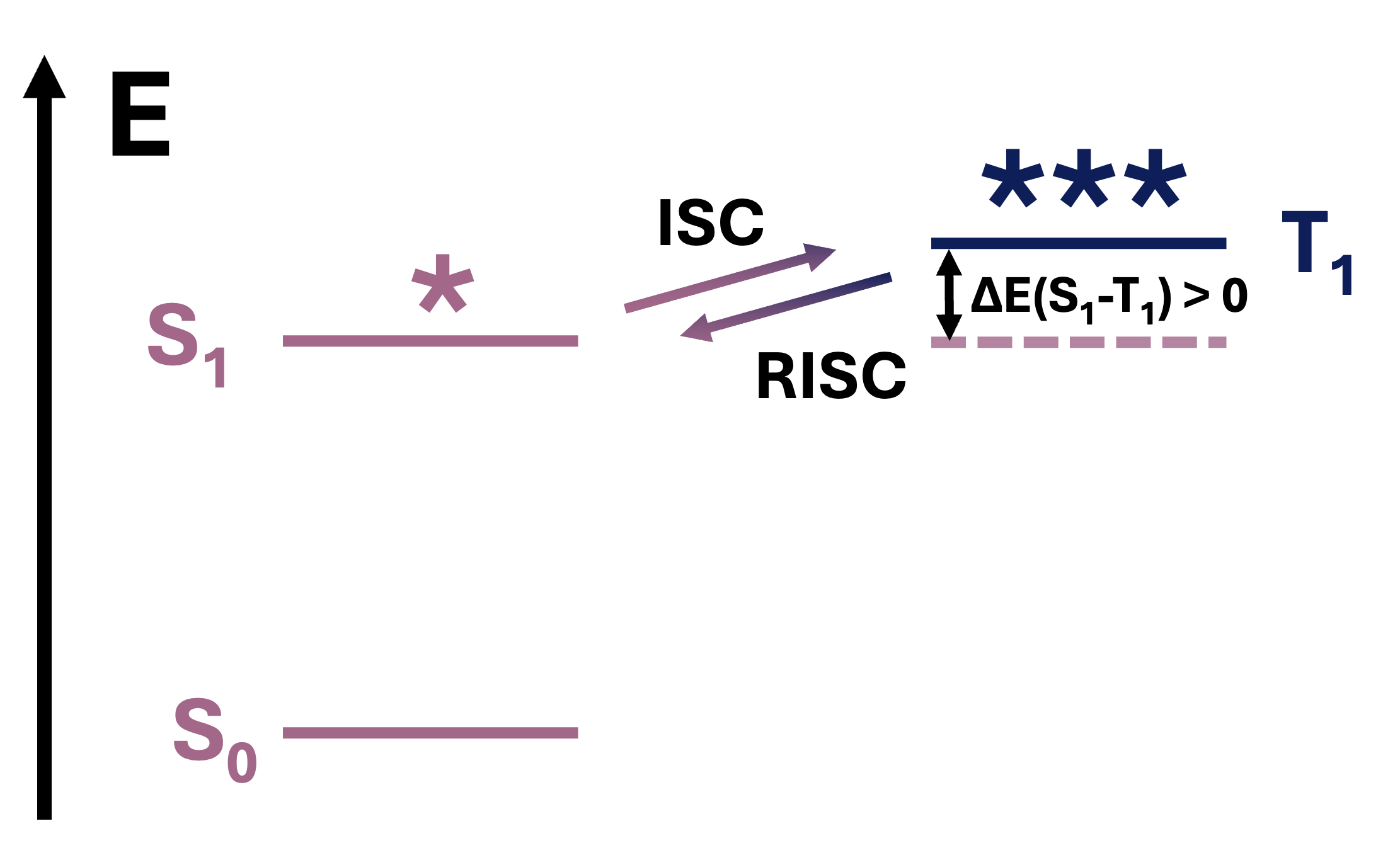}
     \caption{InveST}
     \label{fig::InveST}
     \end{subfigure}
     \hfill
    \caption{Schematic illustration of the SF process (adapted from Ref.~\citenum{smith_singlet_2010}) and the InveST mechanism (adapted from Ref.~\citenum{pollice_organic_2021})}
    \label{fig::SFandInveST}
\end{figure}

The study of singlet fission in molecules is largely motivated by the desire to develop photovoltaics with increased efficiency.
The overall efficiency in solar cells is limited to around $30\%$ (Shockley-Queisser limit\cite{shockley_detailed_1961}) with the assumption that one absorbed photon can yield one electron-hole pair\cite{teichen_microscopic_2012}.
This limit could be circumvented if one photon can generate multiple electron-hole pairs.
For dye-sensitized solar cells it was suggested\cite{hanna_solar_2006} that this could be achieved through the process of singlet fission, where in most discussed cases an organic chromophore in an excited singlet state transfers energy to a neighbouring ground-state chromophore or chromophores, to produce two triplet excited states\cite{smith_singlet_2010} (for a schematic illustration see also Fig.~\ref{fig::SF}).
The efficiency for an ideal solar cell with this process rises to nearly $50\%$\cite{hanna_solar_2006,smith_singlet_2010}.

The inverse design problem to find the right chromophores is complicated by the fact that singlet fission appears to be highly system-specific, with different mechanisms proposed \cite{teichen_microscopic_2012}.
The search for suitable chromophores therefore cannot be limited to finding the optimal electronic structure of the chromophores, but must eventually also include the dynamic evolution \cite{casanova_theoretical_2018} of the two triplet states.
Furthermore, environmental effects may need to be included explicitly, as polar solvents can cause the electronic energy levels of the chromophores to fluctuate in the same order of magnitude as the original isolated chromophores' level splitting \cite{teichen_microscopic_2012}. 

The PPP Hamiltonian has been used in the context of singlet fission in different capacities, and here we can only give a small selection.
One study found design rules based on PPP computed spectra that predict the spectra of acene-based molecules, including oligomers with unusual bonding geometries and hetero-atom substitutions \cite{hele_anticipating_2019}.
Other use cases include in model exact approaches to engineer chromophore properties through substitution with heteroatoms or substituents for a more efficient singlet fission.
Investigated systems range from pyrene  with inter-molecular singlet fission\cite{patra_designing_2024} to diphenylpolyenes that support endoergic to isoergic singlet fission with increasing chain length\cite{karmakar_low-lying_2022}.

Another focus is on the adaption of the PPP model to include additional physical effects such as electron-phonon interactions in polyenes that affect the singlet fission process \cite{valentine_higher-energy_2020}.
The resulting Hamiltonian is called the PPP-Peierls (PPPP) model, and it accounts for the larger flexibility in polyenes compared to rigid structures such as solids or even acenes.
The PPP model was also adapted to include an intermolecular interaction component between two different chromophores, where only a minimal enhancement in performance due to singlet fission was found for a pentacene-C$_{60}$ solar cell\cite{aryanpour_does_2013}.

\subsection{Inverted singlet-triplet energy gap}
\label{subsection::InveST}

Another very challenging inverse design problem is the discovery of more efficient organic light-emitting diode (OLED) materials.
There are several factors that limit the internal quantum efficiency (IQE), which defined as the ratio between the number of generated photons to the number of injected electrons\cite{wang_fundamentals_2019}.
While singlet excitons in OLEDs are fluorescent, triplet excitons are normally non-emissive and decay non-radiatively to the ground state\cite{de_silva_inverted_2019}.
Additionally, the desired singlet excitons are only generated in a 1:3 ratio relative to the triplet excitons, due to the spin statistics of recombining charge carriers\cite{aizawa_delayed_2022}.

The final impediment is embodied in Hund's multiplicity rule that places the first excited singlet state $S_1$ of an organic closed-shell molecule higher in energy than the corresponding first excited triplet state $T_1$.
Defining the singlet-triplet energy gap as $\Delta E_{ST}=S_1-T_1$, molecules that obey Hund's rule therefore have a positive value for $\Delta E_{ST}$.
This indicates that energetically population transfers from the singlet to the triplet state is favoured in a process called intersystem crossing (ISC).

The focus in the development of new OLED materials has therefore been to address these limitations through molecular engineering. The efficiency in OLEDs has been improved by introducing phosphorescent emitters that make the triplet state bright as well as by reducing $\Delta E_{ST}$, so that thermally activated delayed fluorescence (TADF), due to a reverse ISC (RISC), becomes possible.

The theoretical discovery of two stable organic molecules with an inverted singlet-triplet energy gap\cite{ehrmaier_singlettriplet_2019, de_silva_inverted_2019} (InveST), where $\Delta E_{ST} < 0$, marked an important achievement, that has significant potential to improve OLED efficiency.
For InveST molecular systems, the triplet excitons avoid thermalization and are instead converted through an energetically favourable RISC to the fluorescent singlet exciton\cite{anwer_recent_2025}.
A schematic illustration of the InveST mechanism is shown in Fig.~\ref{fig::InveST}.

Finding molecules with InveST properties has been challenging, which is somewhat unsurprising given the dearth of examples of Hund's rule violations either experimentally or theoretically before 2019\cite{pollice_organic_2021}.
There has been a large effort to theoretically predict new InveST molecules and to understand the underlying process to create useful design rules\cite{perezjimenez_role_2025}.
Experimentally, there has also been confirmation for some InveST molecules \cite{ehrmaier_singlettriplet_2019, aizawa_delayed_2022, wilson_spectroscopic_2024, blasco_experimental_2024,kusakabe_inverted_2024}.
A great challenge in treating InveST molecules theoretically has been the inadequacy of uncorrelated excited-state methods and linear response TD-DFT to describe the InveST phenomenon\cite{de_silva_inverted_2019,sobolewski_are_2021}.

It has been shown\cite{kollmar_violation_1978,koseki_violation_1985,borden_violations_1994} that the inclusion of accurate electron correlation is crucial to reproduce inverted gaps.
This is due to one of the characteristic features in InveST molecules: a minimal exchange integral between the frontier orbitals (HOMO and LUMO), that in turn elevates the importance of the spin-polarization \cite{kollmar_violation_1978}, an otherwise smaller effect.
Spin-polarization can then even reverse the sign of $\Delta E_{ST}$.
To describe spin-polarization effects, electron correlation with at least double excitations from an HF reference is necessary\cite{sobolewski_are_2021}. 

The inverse design problem for InveST molecules is a very active research field, where theoretical efforts have taken a prominent role and are complemented by experiment\cite{perezjimenez_role_2025}.
Here again, we will focus on just the PPP Hamiltonian, its specific benefits for the study of InveST molecules and how it has been applied so far.

Some computational challenges, such as the need for higher-order electron correlation treatment, can be more easily overcome with the PPP Hamiltonian.
Furthermore, even the smallest proposed InveST systems are at the edge of what standard correlated \textit{ab\,initio} methods can reach \cite{pollice_organic_2021}.
The PPP Hamiltonian can therefore help to make larger systems accessible and reveal general trends\cite{bedogni_singlettriplet_2024}.
The absolute value of $\Delta E_{ST}$ can be rather small \cite{dreuw_inverted_2023} and is in the same order as the error of the computational methods\cite{perezjimenez_role_2025}.
Chemical accuracy is therefore required to determine the sign of $\Delta E_{ST}$ with confidence, and a PPP model exact approach could help identify critical effects such as environment, correlation contribution etc.
A very early PPP Hamiltonian study, predating the renewed interest in InveST molecules by decades, found molecules such as propalene, pentalene and heptalene in violation of Hund's rule for $D_{2h}$ symmetry, but in agreement with it for a relaxed $C_{2h}$ symmetry\cite{koseki_violation_1985}.

The PPP Hamiltonian has been discussed in the context of high-throughput screening for InveST molecules as a scoring function, to filter out the most promising InveST candidates, that can be investigated more closely by higher-level \textit{ab\,initio} methods\cite{jorner_ultrafast_2024}.
One of the attractive features of the PPP Hamiltonian as a scoring function, is the cheap and simple description relative to other approaches, while still capturing the most important physics\cite{jorner_ultrafast_2024}.
In that case, the electronic structure was solved with configuration interaction singles and perturbative double excitations for a very efficient virtual screening.
The authors mention two main concerns with their findings: the inability to correctly predict oscillator strengths and restriction to $\pi$-electron transitions.
It seems to be an open challenge to strike the balance between a cheap scoring function on the one hand and a sufficiently accurate description, potentially requiring more costly inclusion of correlation interactions.

For triangulene systems with the PPP Hamiltonian, it was shown that triple excitations of the HF reference are non-negligible for the description of the excited states when the transition energy is compared with model exact calculations\cite{bedogni_shining_2024}.
Further work by some of the same authors found that within the PPP model space, a negative $\Delta E_{ST}$ can be explained by a network of alternating electron-donor and acceptor groups in the molecular rim, rather than the triangular molecular structure itself\cite{bedogni_singlettriplet_2024}.

\section{PPP tomorrow}
\label{sec::OutlookQC}
As detailed above, the PPP Hamiltonian can be viewed as a “minimum viable parametrization of conjugated chemistry”, with a long and influential history in the development of new theoretical methods.
It is 'minimal' in that it offers a highly reduced representation of the full electronic structure problem, explicitly treating only the $\pi$-electrons, yet still providing an insightful description for many molecules of chemical or technological interest.
Over the past seven decades, the PPP model was able to provide insights into extended systems that were otherwise unreachable, and it has propelled the development of novel computational methods in quantum chemistry.

As the PPP model is well established nowadays and many smaller systems have been calculated accurately with existing methods, it can be used as a valuable reference to benchmark novel methods.
While the PPP approach is rooted in the same molecular orbital framework as most \textit{ab initio} methods, it replaces the cumbersome integral evaluation with a more lightweight parametrization of the electronic interactions.
Due to this model character, it is easy to tune the parameters and explore specific regimes directly, such as the weak and strong coupling limits, which allows for a straightforward testing of new computational approaches across different correlation regimes.
There has also been substantial progress in clarifying the scope and limits of the model itself, and different parametrization techniques have been developed to maximize the predictive power of the PPP model.

Meanwhile, quantum computing is emerging as a promising alternative to mitigate the unfavourable scaling of traditional quantum chemistry methods by leveraging quantum mechanical principles to represent and manipulate quantum states more efficiently than classical methods. 
While current noisy intermediate-scale quantum (NISQ)-era devices lack the error correction and scalability required for chemically accurate simulations of large systems, rapid hardware progress suggests that the first fault-tolerant quantum computations for chemistry are soon within reach\cite{google_quantum_ai_and_collaborators_quantum_2025}.
However, these early fault-tolerant quantum computers will offer only very limited computational resources, thus requiring reduced problem descriptions and optimal utilization of the available resources to obtain meaningful results\cite{alexeev_perspective_2025}.

Here, we propose the PPP Hamiltonian as an ideal candidate for insightful calculations in such a resource-constrained compute environment, since it can be considered the minimum viable model of organic molecules that still captures the essential chemistry.
The $\pi$-electron approximation of the PPP model allows for a drastic reduction of the Hilbert space size and requires significantly fewer resources compared to an \textit{ab initio} calculation, even when considering only a minimal basis set.
To illustrate this on a simple example, we compare an \textit{ab initio} treatment of benzene with the corresponding PPP model Hamiltonian.
For the \textit{ab initio} description, a minimal STO-3G basis set for the 6 carbon and 6 hydrogen atoms results in a total of 72 spin orbitals, while the PPP Hamiltonian only requires two spin orbitals per carbon atom, thus describing benzene with a total of 12 spin orbitals.
For quantum computing applications, the reduction in the number of spin orbitals translates directly into an equivalent reduction of the number of qubits needed to represent the system.
This makes PPP a prime candidate for calculations on early quantum computers that possess only a limited number of (logical) qubits.

\begin{figure}[htb]
    \centering
    \includegraphics[width=0.9\textwidth]{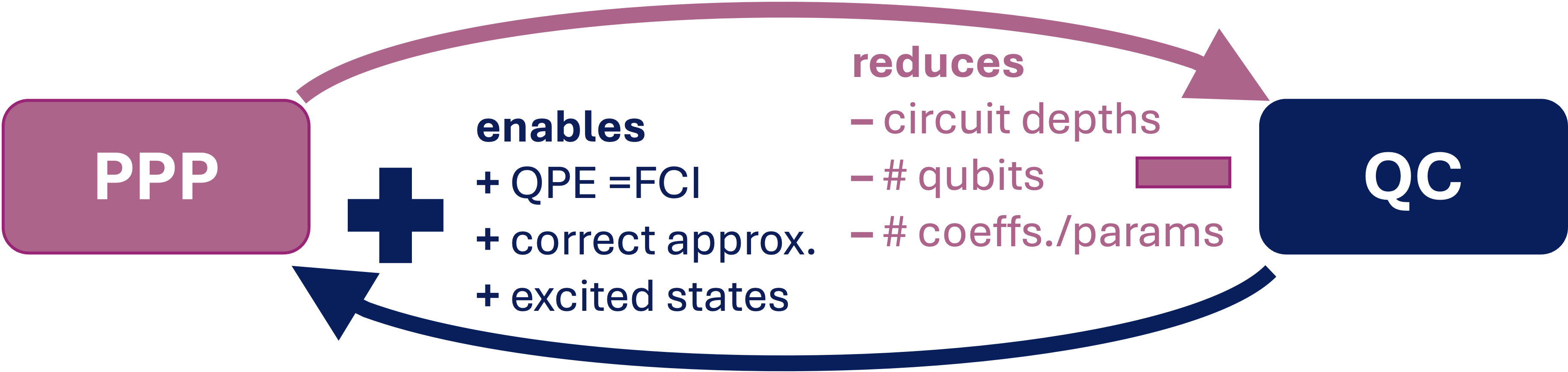}
    \caption{Scheme illustrating the mutual benefits of employing the PPP model in quantum computing applications for chemistry.}
    \label{fig::PPPanQC}
\end{figure}

Another advantage of the PPP model can be found in the ZDO approximation, which leads to a very sparse Hamiltonian matrix.
In the atomic orbital representation that we have used in equation~\eqref{eq:PPP} for the PPP Hamiltonian, all terms are diagonal, except for the nearest-neighbour hopping terms $t_{ij}$.
The same is true for the real-space valence-bond description of the PPP Hamiltonian \cite{soos_valence-bond_1984}, whereas the molecular orbital representation of the PPP Hamiltonian is less sparse and has more terms \cite{mukhopadhyay_neutral_2008, bedogni_singlettriplet_2024}.
The AO representation of the PPP model will have a formal scaling of $O(N^2)$ in the number of Hamiltonian terms without any further screening, where N is the number of orbitals. 
In contrast, the MO representation of the PPP Hamiltonian\cite{mukhopadhyay_neutral_2008} and general \textit{ab initio} Hamiltonians \cite{helgaker_molecular_2004} formally scale as $O(N^4)$.
The actual number of terms for the MO PPP Hamiltonian will still be greatly lower than an \textit{ab\,initio} Hamiltonian due to the neglect of three- and four-site two-electron integrals in the ZDO approximation.
For quantum computing applications, sparser Hamiltonians require fewer terms to be encoded on the device, thereby reducing the number of gate operations needed to represent a given Hamiltonian in a quantum circuit.
Another potential advantage of the PPP model in that context is that many interaction parameters will be identical, especially for idealized molecular structures, such that these terms can be grouped and implemented more efficiently\cite{bay-smidt_fault-tolerant_2025,gidney_halving_2018,kivlichan_improved_2020}.
The advantages of the PPP model for quantum computing are summarized schematically in Fig.~\ref{fig::PPPanQC}.

While the PPP Hamiltonian is ideally suited as an early test bed for developing quantum computing applications, the reverse is also true, as illustrated in Fig.~\ref{fig::PPPanQC}.
The PPP model, conceived in a resource-constrained environment and therefore often used in more approximate treatments of systems, benefits greatly when more correlation contributions can be considered, ideally solving the system exactly.
This is particularly true when calculating excited-state properties, as these may necessitate a multi-reference treatment of the wave function while capturing multiple, potentially close-by states.
Furthermore, some approximations discussed before for the PPP model, are only truly valid when considerable electron correlation is included.
The pursuit of exact solutions has been prohibitive beyond modest system sizes within the classical computing paradigm due to the exponential scaling of the underlying problem.

Here, the quantum computing equivalent of an FCI calculation, the quantum phase estimation algorithm (QPE), offers a more favourable polynomial scaling compared to the exponentially scaling classical FCI computation.
The minimal description of the $\pi$-electron PPP Hamiltonian provides an additional advantage to reach larger systems efficiently with limited quantum resources, especially in the calculation of spectroscopic properties that are classically unfeasible.
The combined quantum semi-empirical approach (QPE+PPP) will also be more efficient than a minimal quantum-\textit{ab\,initio} approach (QPE+STO-3G, as discussed before for the benzene molecule).
Strong correlation effects will be better and more efficiently described in the minimal semi-empirical calculation than the minimal \textit{ab\,initio} calculation, due to their implicit inclusion in the parametrization\cite{segal_semiempirical_1977-1,martin_ab_1994-1}.
Moreover, the included correlation interaction in the QPE calculation will ensure that the approximations of the PPP Hamiltonian ($\pi$-electron, ZDO, transferability of the semi-empirical parameters) are truly valid analogous to the classical model exact studies.

In the same spirit as in the classical case, the PPP model need not only be applied in FCI-type calculations such as QPE, but can also serve as a benchmark model to validate more approximate quantum algorithms.
Similarly to its use in the development of approximate correlated classical methods, particularly for CC-variants and DMRG applications in chemistry, the PPP Hamiltonian could be used as a resource-efficient test system for early fault-tolerant algorithms that only capture a limited degree of correlation, such as quantum subspace methods, filtering techniques, and statistical approaches\cite{katabarwa_early_2024}.

For these reasons, we conjecture that the PPP Hamiltonian is ideally suited for early fault-tolerant quantum computing applications and vice versa.
Within this framework, different avenues and possibly new tradeoffs might present itself that were traditionally not considered.
As an example, we return to the previously mentioned choice of representing the PPP model in three different bases (AO/MO/VB).
It is not clear if the MO representation, which is predominant in classical applications, is also the optimal choice for a quantum computer.
For the PPP Hamiltonian, the AO representation offers the most compact representation and also the least resource demands.
The traditional drawback of the AO representation against the two other approaches, MO and VB, has been the difficulty to truncate this basis.
Therefore, AO representations of the PPP Hamiltonian commonly necessitate an FCI calculation, something that comes naturally with the QPE algorithm.

The largest classical FCI calculation to date was for the propane molecule, \ce{C3H8}, with a minimal STO-3G basis which required the description of 26 electrons in 23 spatial orbitals (equivalent to 46 spin orbitals)\cite{gao_distributed_2024}.
This limitation of classical FCI to a double-digit number of spin orbitals highlights the potential utility of quantum computers already with a comparable number of logical qubits, particularly when employing model Hamiltonians such as the PPP model.
By pairing QPE with PPP, one can achieve a chemically meaningful yet highly efficient description in terms of qubit resources which requires only very limited circuit depth thanks to the sparsity of the PPP Hamiltonian.
Hence, this combination is a particularly promising candidate for early impactful applications of fault-tolerant quantum computing in chemistry.

\begin{acknowledgement}

This work is supported by the Novo Nordisk Foundation, Grant number NNF22SA0081175, NNF Quantum Computing Programme. This work is partly funded by Innovation Fund Denmark (IFD) under File no. 4356-00009B. We thank Matthew Teynor for helpful comments on the manuscript.

\end{acknowledgement}

%
%

\bibliography{MyLibrary}

\end{document}